\documentclass[conference]{IEEEtran}

\usepackage{amsmath}
\usepackage{algorithm}
\usepackage{algpseudocode}
\usepackage{graphicx}

\usepackage{subfloat}
\usepackage{subfig}

\usepackage[utf8]{inputenc}
\usepackage[T1]{fontenc}
\usepackage{lmodern}
\usepackage{lmodern}

\usepackage{array}    

\newcommand\ceil[1]{\lceil#1\rceil}

\begin{document}

\title{An  Efficient Keyless Fragmentation Algorithm for Data Protection}

\author{\IEEEauthorblockN{Katarzyna Kapusta, Gerard Memmi, and Hassan Noura }
\IEEEauthorblockA{Télécom ParisTech, Université Paris-Saclay\\
23, avenue d'Italie, \\
75013 Paris, France \\
Email: \{firstname.lastname\}@telecom-paristech.fr}
}

\maketitle

\begin{abstract}
The family of Information Dispersal Algorithms is applied to distributed systems for secure and reliable storage and transmission. In comparison with perfect secret sharing it achieves a significantly smaller memory overhead and better performance, but provides only incremental confidentiality. 
Therefore, even if it is not possible to explicitly reconstruct data from less than the required amount of fragments, it is still possible to deduce some information about the nature of data by looking at preserved data patterns inside a fragment.  The idea behind this paper is to provide a lightweight  data fragmentation scheme, that would combine the space efficiency and simplicity that could be find in Information Dispersal Algorithms with a computational level of data confidentiality. 

\end{abstract}

\section{Introduction}

Fragmenting data and dispersing them over different physical locations through several transmission paths slows down an attacker from obtaining the totality of the original data. Data are usually transformed into fragments by the use of secret sharing~\cite{Shamir:1979:SS:359168.359176}, information dispersal algorithms~\cite{rabin:1989}, or data shredding~\cite{systemx,Fabre1994}. 

The choice of the most appropriate fragmentation method depends on the particularity of a given use case. A user has to balance between memory use, performance and desired confidentiality level. Fragments obtained with perfect secret sharing are highly secure, but the technique is slow and very costly in memory. Information dispersal is resilient and relatively fast, but not secure. Combining symmetric encryption with fragmentation is secure and easily scalable, but in some circumstances may be less efficient than information dispersal~\cite{Resch:2011}.

A fragmentation technique for long-term archival storage of large data will usually differ from the one applied to disperse small data packets. In the last case, the choice would be to pick some lightweight fragmentation technique, like information dispersal or a fast computational secret sharing scheme, rather than to apply more complex mechanisms~\cite{Katti:slicing,Parakh:onlinedatastorage}.

The idea of an Information Dispersal Algorithm(IDA) was first introduced by Rabin in the late 80s~\cite{rabin:1989}. Broadly, the algorithm multiplies data chunks by a matrix in order to obtain fragments that are a linear combination of the chunks' data and matrix elements. The recovery is only possible when a certain amount of these fragments is being gathered. Rabin's IDA has several advantages: it adds resilience to data, produces almost no storage overhead and uses simple arithmetic operations. Although, the scheme guarantees only incremental confidentiality. An eavesdropper knowing the dispersal matrix can verify if a fragment has a predetermined value. Moreover, such attacker can guess the content of missing fragments when data have recognizable patterns. Despite of such problems, IDAs are still taken into consideration in the context of data protection, as the obstacle of not being able to explicitly reconstruct initial data from less than required amount of fragments may be sufficient in some scenarios~\cite{secret:shares}. Such scenarios include all use cases were memory or performance overhead caused by secret sharing or cryptographic operations may be a burden and data protection requirements are not too high.
An example could be a fragmented database~\cite{cuppens, bib:vimercati} where we would like to quickly retrieve records or a multipath transmission where we would like to defragment only the next destination address~\cite{Katti:slicing}.

A new scheme situated between computational secret sharing and information dispersal algorithms was recently introduced in a three pages
poster~\cite{Kapusta:2016}. Its complexity and storage overhead are comparable to the one of the IDAs, but it provides higher level of data confidentiality. This paper significantly extends the previous proposal. It contains detailed descriptions, as well as presents more experimental security and performance analyses realized on industrial data.

This paper is organized as follows. Section~\ref{sec:notation} summarizes the notation used. Section~\ref{sec:contribution} contains a description of our contribution. Section~\ref{sec:relevant} presents related works. Section~\ref{sec:algorithm} describes the scheme. Section~\ref{sec:security} presents its empirical security evaluation. Section~\ref{sec:cryptanalysis-discussion} contains a cryptanalysis discussion. Section~\ref{sec:performance} shows performance results. An insight into future works ends the paper.

\section{Notation}
\label{sec:notation}
In order to unify descriptions, we are introducing the notation presented in Table~\ref{tab:notation}.

\begin{table}[h]
\caption{Notations}
\label{tab:notation}
\centering
\def\arraystretch{1.5}
\begin{tabular}{|l|l|}
\hline
    $d$ & initial data \\ [0.02cm] \hline
   $d_{size}$ & size of the initial data \\  [0.02cm]  \hline
   data chunk  & one or more consecutive bytes of original data \\ [0.02cm] \hline
   data share & an encoded data chunk \\ [0.02cm] \hline
   fragment  & a final data fragment composed of data shares \\ & and stored in one location \\ [0.02cm] \hline
   $k$ & number of fragments required for data recovery \\  [0.02cm]  \hline
   $l$ & number of data chunks inside original data \\  [0.02cm]  \hline
   $n$ & total number of fragments \\  [0.02cm]  \hline
   $DCS$   & Data Chunk Set, a set $k$ of data chunks  \\ [0.02cm] \hline
   $DSS$   & Data Share Set, a set $k$ of data shares  \\ [0.02cm] \hline
   $S$ & seed, a set of $k$ pseudorandom values \\  [0.02cm]
  \hline
\end{tabular} 
\end{table}

\section{Our contribution}
\label{sec:contribution}

The proposed algorithm fragments initial data $d$ of size $d_{size}$ into $n$ fragments of a size close to $\frac{d_{size}}{k}$, any $k$ of which are needed for data recovery. Initial data are processed by sets of $k$ chunks. Data encoding is not based on a matrix multiplication, but on a modification of Shamir's secret sharing scheme~\cite{Shamir:1979:SS:359168.359176} and depends on the encoding results of the previously processed $k$ data chunks. A pseudo-random seed is used as the first set of data chunks and dispersed within the data. 

The main benefits of such scheme are: 
\begin{itemize}
\item No symmetric key encryption is applied, the processing makes use only of simple operations.
\item Data patterns are not being preserved inside fragments and the content verification is not straightforward(in contrast to IDAs).
\item Partial data defragmentation is possible(in contrast to schemes based on the all-or-nothing transform).
\end{itemize}

\section{Related works}
\label{sec:relevant}

This section presents most relevant works from the domains of secret sharing and information dispersal.

\subsection{Information Dispersal Algorithms}
\label{subs:ida}  

An Information Dispersal Algorithm~\cite{rabin:1989} divides data $d$ into $n$ fragments of size $\frac{d_{size}}{k}$ each, so that any $k$ fragments suffice for reconstruction. More precisely, $n$ data fragments are obtained by multiplying initial data by a $k \times n$ nonsingular generator matrix . Recovery consists in multiplying any $k$ fragments by the inverse of a $k \times k$ matrix built from $k$ rows of the generator matrix. Information dispersal adds redundancy to data and does not produce storage overhead. In~\cite{Li:ida}, Li analyzed the confidentiality of IDAs. For instance, Rabin's  IDA proposal was evaluated to have strong confidentiality, as the original data cannot be explicitly reconstructed from fewer than the $k$ required fragments. However, even if it is not possible to directly recover the data, some information about the content of the initial data is leaked. Indeed, data patterns are preserved inside the fragments when the same matrix is reused to encode different data chunks. A similar problem occurs while using the Electronic Code Book block cipher mode for block cipher symmetric encryption~\cite{Dworkin:2001:SER:2206247}.


\subsection{Shamir's secret sharing}
\label{subs:shamir}

Shamir's perfect secret sharing scheme~\cite{Shamir:1979:SS:359168.359176} takes as input data $d$ and fragments them into $n$ fragments $f_1,...,f_n$, of which at least $k$ are needed for data recovery. The algorithm is based on the fact that given $k$ unequal points $x_1,...,x_k$ and arbitrary values $f_1,...,f_k$ there is at most one polynomial $y(x)$ of degree less or equal to $k-1$ such that $y(x_i)=f_i, i=1,...,k$. The algorithm provides with the highest level of confidentiality, but has quadratic complexity in function of $k$ and produces fragments of size equal to the initial data. Therefore, it is usually applied for protection of smaller data like encryption keys. In such a use case, drawbacks of the scheme are acceptable, but for larger data they may be a major obstacle. 

\subsection{Secret Sharing Made Short}
\label{subs:ssms}

Krawczyk's Secret Sharing Made Short (SSMS)~\cite{Krawczyk:1993:SSM:646758.705700} combines symmetric encryption with perfect secret sharing for protection of larger data. Data $d$ are encrypted using a symmetric encryption algorithm, then fragmented using an Information Dispersal Algorithm. The encryption key is fragmented using a perfect secret sharing scheme and dispersed within data fragments. In consequence, the solution does not require explicit key management. The storage overhead does not depend on data size, but is equal to the size of the key per data fragment. The performance of the SSMS technique depends on the details of the chosen encryption and IDA techniques.

\subsection{AONT-RS}
\label{subs:aont-rs}

The AONT-RS technique~\cite{Resch:2011} is similar to SSMS, as it combines symmetric encryption with data dispersal. It applies an all-or-nothing transform(AONT)~\cite{Rivest:97} to create $k$ fragments: encrypted data are divided into $k-1$ fragments and an additional fragment is generated by {\it xor}-ing hashes of these data fragments with the key used for encryption. Additional $n-k$ fragments are produced using a systematic Reed-Solomon error correction code. Data integrity is ensured by the use of a canary that is dispersed within the fragments.


\subsection{Parakh's scheme}
\label{subs:parakh}

A steganographic threshold scheme presented in \cite{Parakh:2011335} transforms $k$ data chunks into $n$ data fragments using a single polynomial of degree $k-1$. The size of produced fragments depends on the value of $k$: it decreases while the number of data chunks and the degree of the polynomial are growing. 


\section{Fragmentation algorithm}
\label{sec:algorithm}

This section describes in details the processing core of the proposed fragmentation scheme. Algorithm \ref{alg:fragmentation} presents the steps of the fragmentation procedure. Data defragmentation is not presented in the form of an algorithm, as it is basically a direct inverse of fragmentation.

\textbf{Data processing flow} Initial data $d$ are treated as a concatenation of $l$ data chunks. These $l$ data chunks are encoded one by one into
$l$ data shares. Further, $n$ fragments are constructed from data shares in a way that $k$ fragments are sufficient for the recovery of $d$. For a more convenient processing, data chunks are regrouped into Data Chunk Sets of $k$ elements, where $DCS_i(j)$ is the $jth$ data chunk in the set $i$. Initial data $d$ may be then presented as a concatenation of $DCSs$: ${DCS_1,...,DCS_{m}}$ ($m=\ceil{\frac{l}{k}}$). Similarly, data shares are regrouped into Data Share Sets of $k$ elements, so the result of encoding is a concatenation of $DSSs$: ${DSS_1,...,DSS_{m}}$. At the end, data shares are distributed to $k$ final fragments and $n-k$ redundant fragments are added.

\begin{figure}[!h]
\centering
\includegraphics[width=0.95\linewidth]{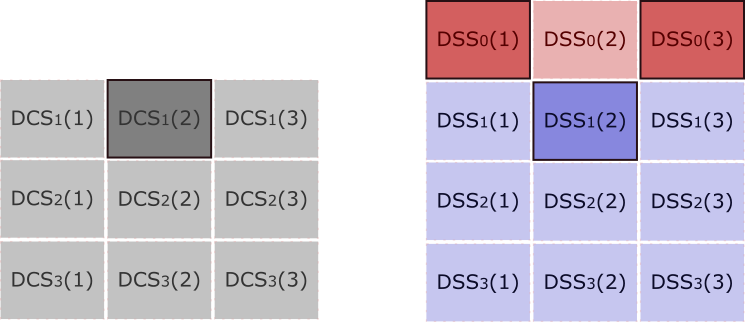}
\caption{{\it Example for $k=3$: data chunks are transformed into data shares. Encoding of $DCS_i$ is based on $DSS_{i-1}$. A pseudorandom seed  serves as $DSS_0$.  $DSS_1(2)$ comes from a transformation of $DCS_1(2)$.  $DSS_0(1)$ and $DSS_0(3)$ were used as coefficients of the encoding polynomial. }}%
\label{fig:dcs}%
\end{figure}

\textbf{Encoding} Data processing is done in a Shamir's like fashion: each data chunk is encoded as a constant term of a polynomial of degree $k-1$. 
More precisely, a data chunk is transformed into a data share inside the function $Encode$. $Encode$ takes as parameters the value of the data chunk to be encoded, coefficients $Coeffs$ of the polynomial, and a point $x$ at which the polynomial will be evaluated. For each data chunk, $x$ and $Coeffs$ are calculated in function of $DSS_{i-1}$ and $j$. Therefore, to recover a single data chunk, a user has to possess the $DSS_i$ containing the result of $Encode$ for that data chunk, as well as the previous $DSS_{i-1}$. An example of that is shown in Figure \ref{fig:dcs}. Reusing of data shares trades the perfect security of Shamir's scheme for a better performance and a smaller size of fragments: during processing the polynomial is evaluated only at one value. Varying not only the polynomial coefficients, but also the values of $x$, prevents the preservation of data patterns inside encoded fragments. 

\textbf{Seed} The $Encode$ function transforming a data chunk into a data share takes as input $k-1$ values of previously encoded data shares. The first set of data chunks $DCS_1$ does not possess a predecessor. Thus, a seed composed of $k$ pseudorandom values is introduced as $DSS_0$. 

\textbf{Distributing to fragments} After data encoding, data shares are distributed over $k$ final fragments ${f_1,...,f_k}$ inside the $DistributeShares$ function, in a way that a data share $DSS_i(j)$ goes to a fragment $f_j$.

\textbf{Adding redundancy} In a final step $n-k$ redundant fragments $f_{k+1},...,f{n}$ are added by the function $AddRedundancy$ implementing a systematic version of a Reed-Solomon error correction code~\cite{RS}. 

\begin{algorithm}
\caption{Fragmentation procedure}\label{alg:fragmentation}
\label{alg:frag}
\begin{algorithmic}[1]
\State $d={DCS_1,...,DCS_m}$
\State $DSS_0={s_1,...,s_k}$
\For {$i = 1:m$}
\For {$j = 1:k$}
\State $x = j \oplus DSS_{i-1}(j)$
\If {$x == 0$}
 $x =1$
\EndIf
\State $Coeffs= DSS_{i-1} \setminus DSS_{i-1}(j)$
\State $DSS_i(j) = Encode(DCS_i(j), Coeffs, x)$ 
\EndFor
\EndFor
\State ${f_1,...,f_k}=DistributeShares({DSS_0,...,DSS_{m}})$
\State ${f_{k+1},...,f{n}}=AddRedundancy(f_1,...,f_k)$
\end{algorithmic}
\end{algorithm}




\subsection{Main characteristics}

\textbf{Complexity} Data fragmentation into $k$ fragments has linear complexity $O(k)$, as encoding a single data chunk is equal to evaluating a value of a polynomial of degree $k-1$ at a single point.  Processing redundant fragments depends on the implementation of the error correction code. Same for defragmentation. 

\textbf{Parallelization} Defragmentation can strongly benefit from parallelization, as each data chunk is recovered independently from others. Larger data are divided into blocks before applying Algorithm~\ref{alg:fragmentation} to partially parallelize also the fragmentation processing. 

\textbf{Partial defragmentation}  An interesting property of the scheme is its fine-grained granularity during  defragmentation. Indeed, to defragment a single data chunk it is only required to know its position inside a fragment, as well as possess the previous data share set. 


\textbf{Fragment size} The size of produced fragments is close to the space efficient value of $d_{size}/k$, proposed by Rabin. The only data overhead comes from the seed, which is generated at the beginning of the fragmentation procedure and then attached to data fragments. Seed size depends on the chosen size of the data chunks, as seed values are of the same size than data chunks. Therefore, the fragmentation procedure produces a data overhead of size of one data chunk per fragment.



\section{Experimental Security Evaluation}
\label{sec:security}

An experimental analysis of security characteristics of the scheme based on the methodology presented in~\cite{noura} was performed. Tests were adapted to the fragmenting nature of the scheme. For instance, we measured the behavior of several parameters (entropy, correlation coefficient, probability density function) in function of number of fragments $k$.  A secure fragmentation algorithm should ensure high level uniformity and independence of fragmented data. Section~\ref{sec:security:uniformity} and~\ref{sec:security:independence} analyze these two properties. Moreover, in Section~\ref{ssec:seed-sensitivity} we test the sensitivity of the scheme to changes inside the seed.

\begin{figure}[!h]
\centering
\subfloat[][{\it Original data}]{\includegraphics[width=0.3\linewidth]{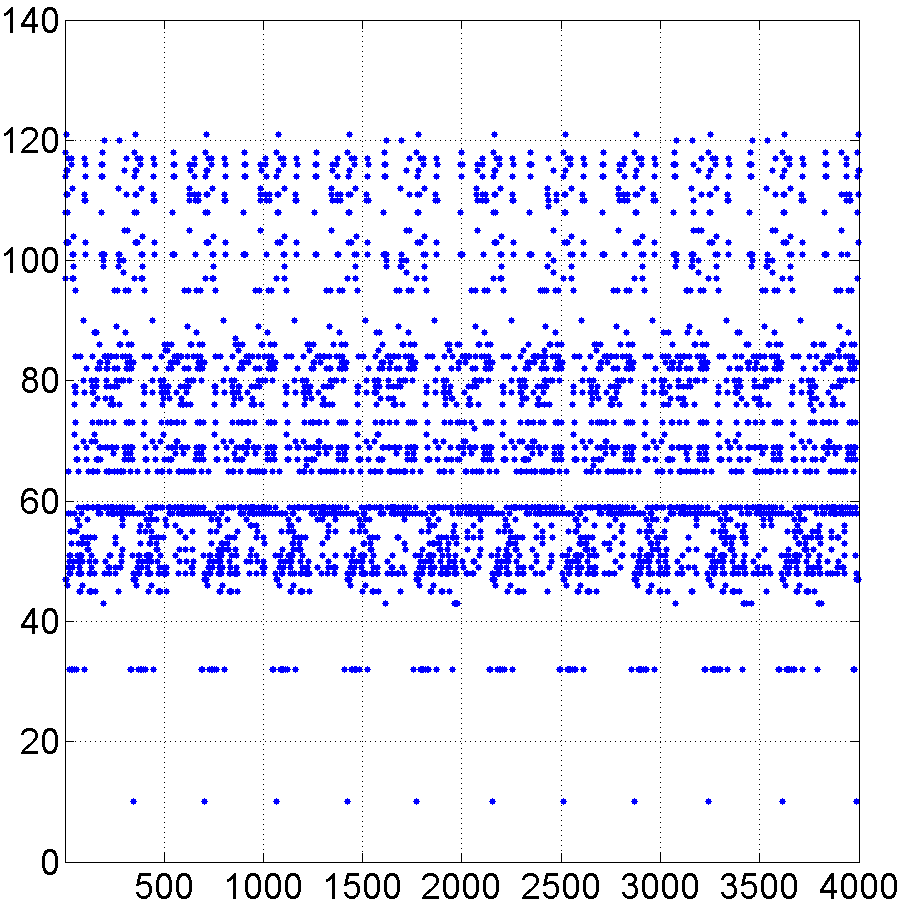} \label{fig:OM}}
\subfloat[][{\it IDA}]{\includegraphics[width=0.3\linewidth]{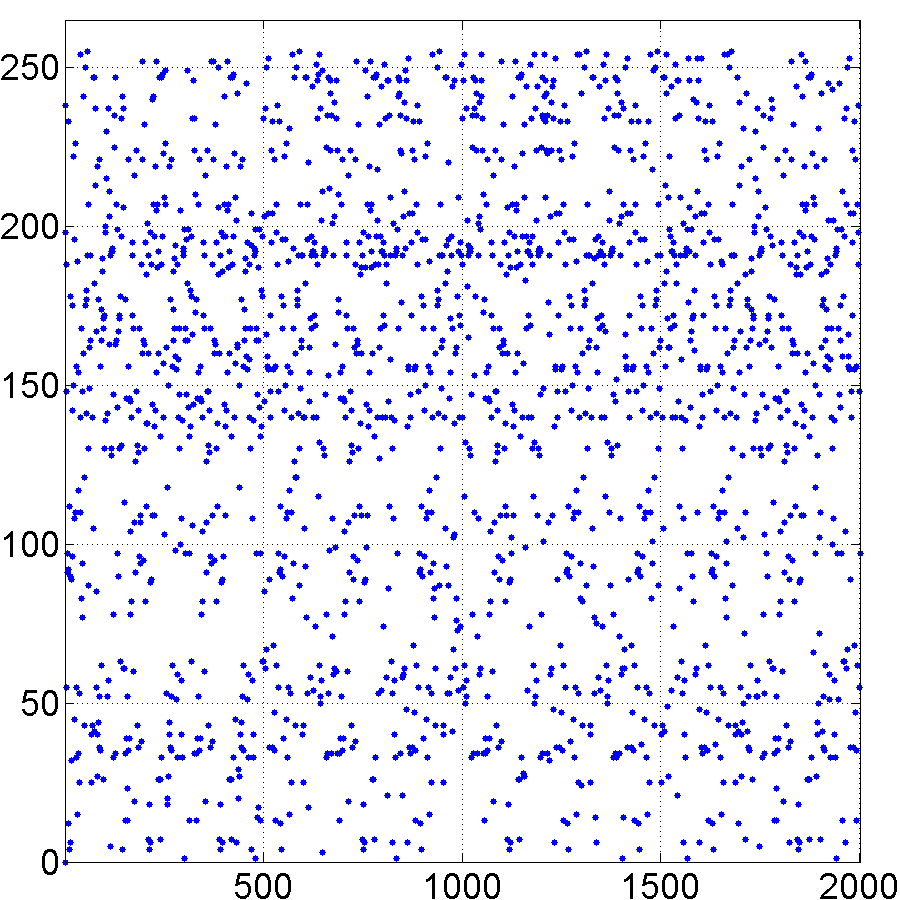}\label{fig:FRAB}} \hspace{0.03in}
\subfloat[][{\it Proposed scheme}]{\includegraphics[width=0.3\linewidth]{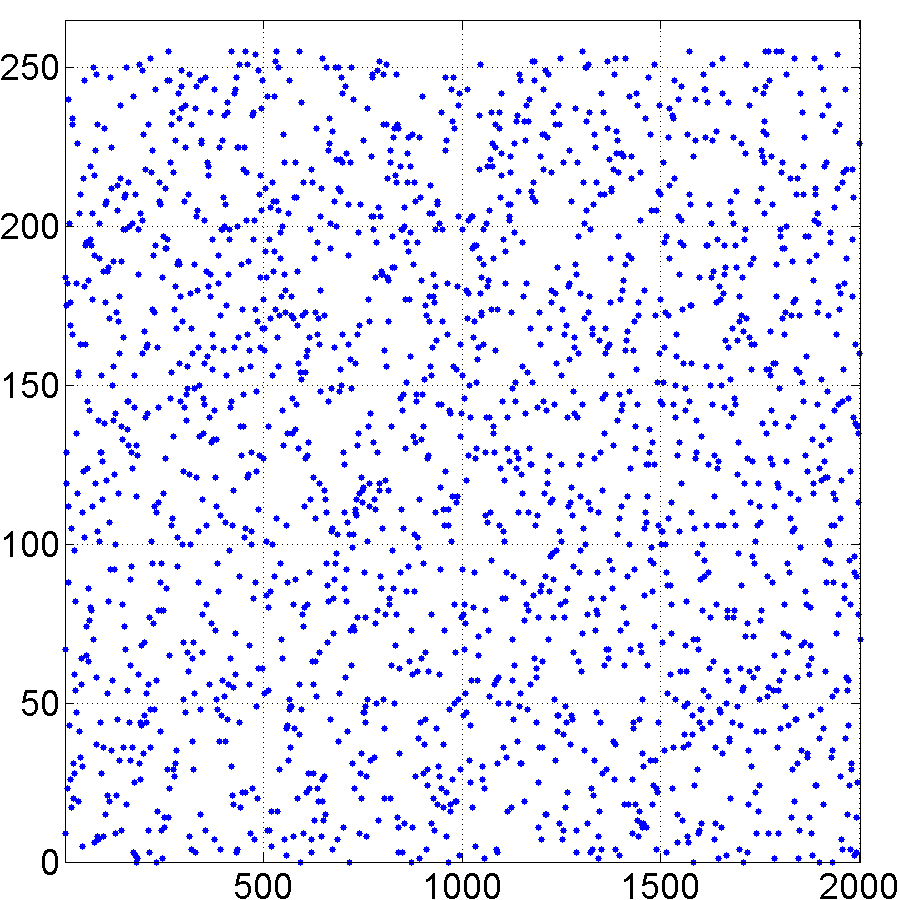} \label{fig:FM}}
\caption{{\it Distribution of a typical textual data sample (a) and distributions of one of its fragments after applying an IDA (b), and after applying the proposed approach (c), for k=2. Data patterns are preserved after the use of an IDA. Fragment (c) contains all possible byte values and does not contain visible data patterns. x-axis shows the byte position inside the sample, y-axis shows the value of the byte at position $x$.}}%
\label{fig:distr}%
\end{figure}

All tests were performed using Matlab environment on textual data samples provided by LaPoste\footnote{http://www.laposte.fr/}. An example of one of such data sample is shown in Figure~\ref{fig:OM}. Its corresponding fragment is presented in Figure~\ref{fig:FM} and compared to the one obtained using an IDA (Figure~\ref{fig:FRAB}). Matlab $rand$ function was used for the generation of the pseudorandom seed.

\vspace{-1.5em}

\vspace{-1em}
\subsection{Uniformity}
\label{sec:security:uniformity}

Encoded fragments should be characterized by high data uniformity, which is an essential property of a scheme resistant against frequency analysis. 
We measure fragments uniformity by visualizing their probability density functions, measuring their entropy, as well as applying the chi-squared test.

\subsubsection{Probability Density Function}

\begin{figure}[!h]
\centering
\includegraphics[width=1\linewidth]{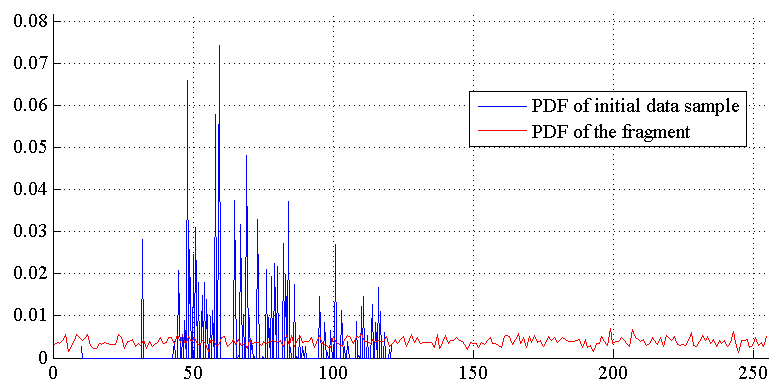}\label{fig:pdff}
\caption{{\it Probability density functions of data sample from Figure~\ref{fig:OM} and its fragment from Figure~\ref{fig:FM}. x-axis shows possible byte values in the sample, y-axis shows the probability of occurrence of a value. }}%
\label{fig:pdf}
\end{figure}

\begin{figure}[!h]
\centering
\includegraphics[width=1\linewidth]{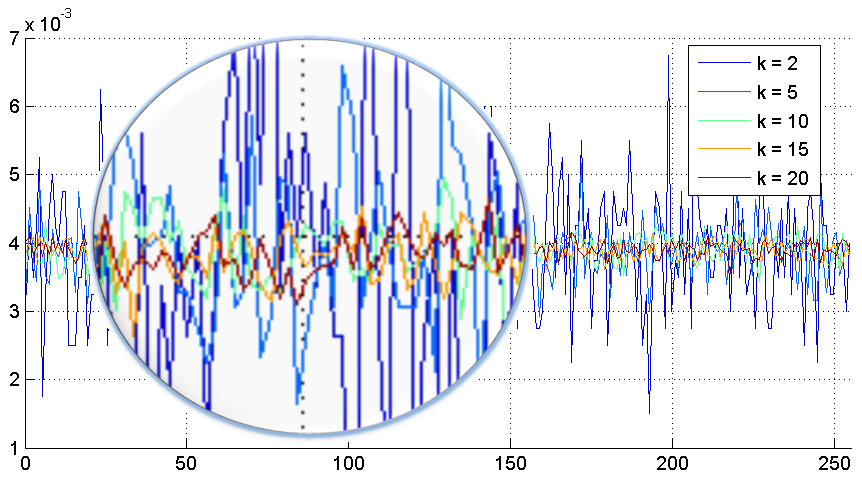}
\caption{{\it Probability density function in function of number of fragments $k$ for different fragments of the data sample from Figure~\ref{fig:OM}. x-axis shows possible byte values in the sample, the y-axis shows the probability of occurrence of a value. $k$ value varies from 2 (dark blue line) to 20 (dark red line). With increasing of $k$ the occurrence probability is getting closer to the one of a uniform distribution (0.039).}}
\label{fig:pdfk}%
\end{figure}

Frequency counts close to a uniform distribution testify data have a good level of mixing. This means that each byte value inside a fragment should have an occurrence probability close to $\frac{1}{v}=0.0039$, where $v$ is the number of possible values (256 for a byte). In Figure~\ref{fig:pdf}, the probability density function (PDF) of a data sample and one of its fragment (for $k=2$) are shown. Results for the fragment are spread over the space and have a distribution close to uniform. In Figure~\ref{fig:pdfk}, the PDF function is also shown, but for different values of $k$ (from 2 to 20). It demonstrates clearly that the occurrence probability of byte values is getting closer to 0.0039 as the value of $k$ is increasing.
\vspace{-1em}

\vspace{-2em}

\subsubsection{Entropy}

\begin{figure}[!h]
\centering
\includegraphics[width=1\linewidth]{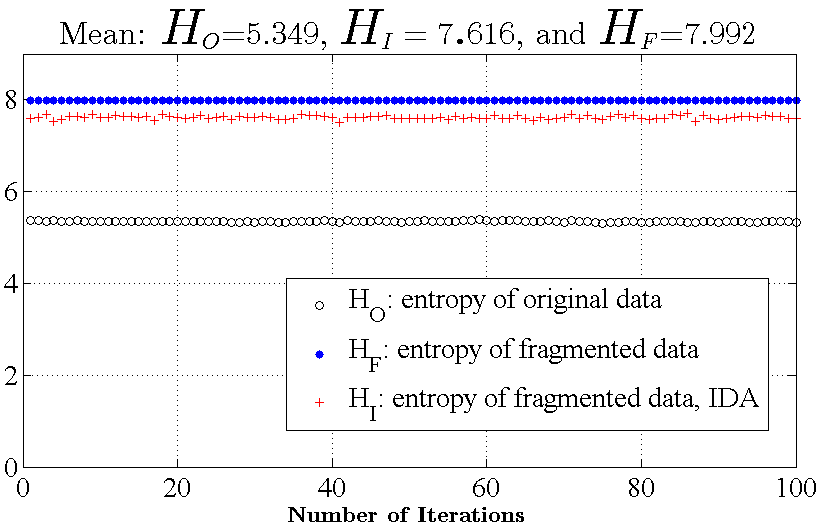}
\caption{{\it Entropy comparison between original and fragmented(with an IDA and with our scheme) data, tested for 100 different data samples, for $k=5$. y-axis shows entropy values (maximum entropy value is equal to 8), the x-axis the number of the comparison test.}}%
\label{fig:entropy}
\end{figure}

Information entropy is a measure of unpredictability of information content~\cite{entropy}. In a good fragmentation scheme the entropy of the fragments should be as high as possible. Figure \ref{fig:entropy} shows entropy variation for all data fragments compared to entropy variation of original data. The chosen data chunk size is equal to one byte, so the maximum entropy value is equal to 8. The average measured value for overall fragments (7.9926) is significantly higher than the one of original data (5.3498) and close to the maximum. This consequently demonstrates that our scheme ensures the uniformity property.

\begin{figure}[!h]
\centering
\includegraphics[width=1\linewidth]{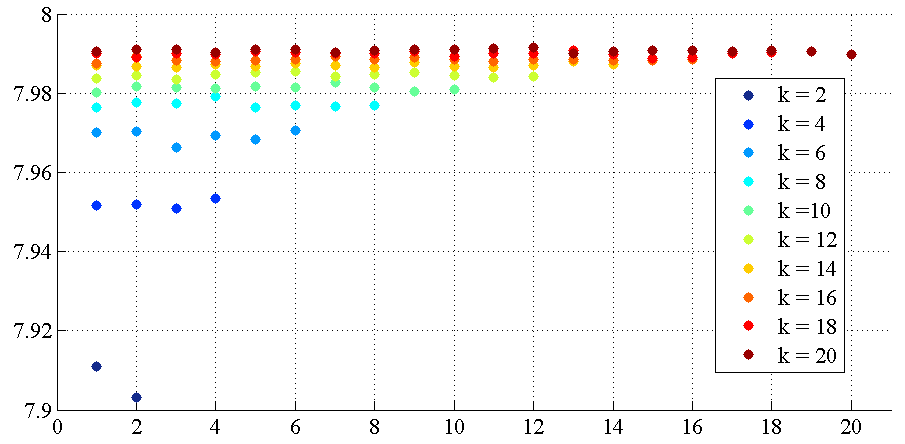}
\caption{{\it Variation of entropy in the function of fragments coming from different fragmentation results, $k$ goes from 2 (dark blue dots) to 20 (dark red dots). x-axis shows the fragment's identifier, y-axis fragment's entropy value.}}
\label{fig:entropyK}
\end{figure}

Furthermore, Figure~\ref{fig:entropyK} shows the entropy level in function of the number of fragments $k$. Each entropy value was obtained as an average of entropy of fragmentation results coming from 10 different data samples.The size of one data fragment was set to 1000 bytes. The result  shows that the entropy is growing with the number of fragments: it starts with a values close to 7.91 for a $k=2$ and achieves 7.99 for values of $k$ close to 20. Moreover, the entropy level is in the same range of values for all $k$ fragments coming from one fragmentation result. This demonstrates that the information is distributed evenly among the fragments.

\vspace{-1em}

\vspace{-1em}

\subsubsection{Chi-squared test}

Uniformity of data inside fragments was validated by applying a chi-squared test~\cite{chi2}. For a significance level of 0.05 the null hypothesis is not rejected and the distribution of the fragment data is uniform if $\chi^2_{test} \leq \chi^2_{theory}(255,0.05)\approx293$. The test was applied on fragmentation results of 15 different data samples for a $k$ going up to 20 and fragment size of 1000 bytes. For all values of $k$, the tests was successful. There was no visible correlation between the number of fragments $k$ and the results of chi-squared test.

\subsection{Independence}
\label{sec:security:independence}

Fragmented data should be greatly different from its original form. To evaluate fragments' independence, we analyze recurrence plots, as well as correlation and bit difference between data and their fragments. 

\subsubsection{Recurrence}

\begin{figure}[!h]
\centering
\subfloat[][{\it Original data}]{\includegraphics[width=0.3\linewidth]{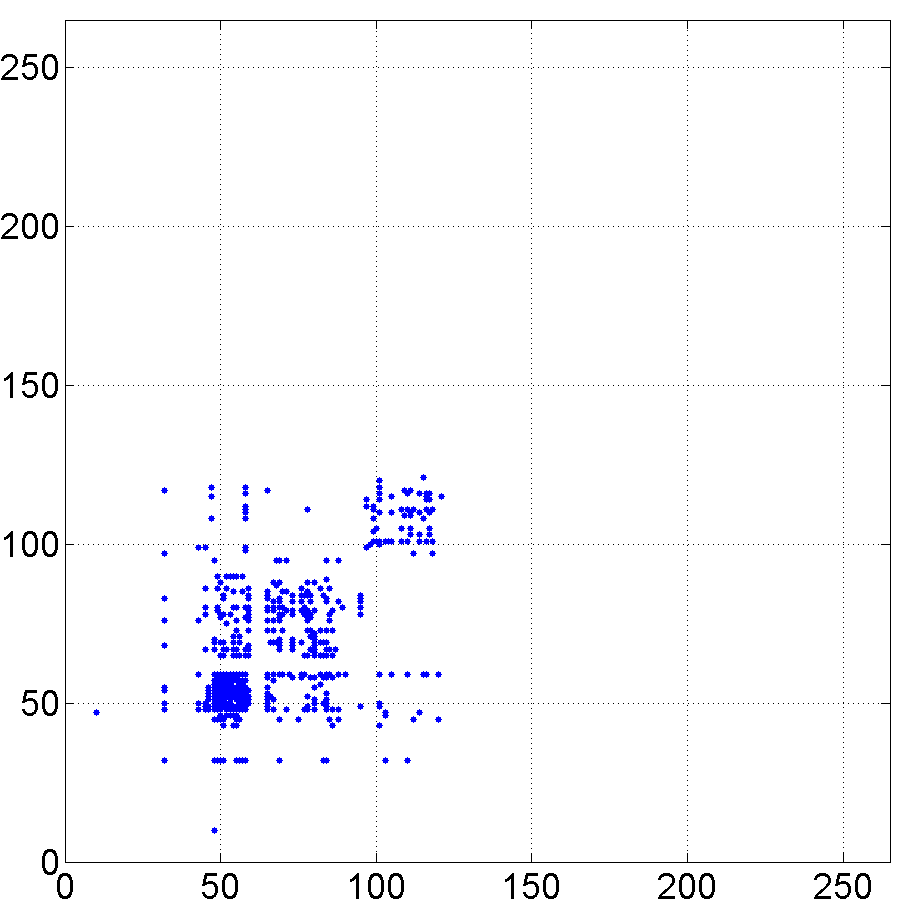}\label{fig:reco}}%
\subfloat[][{\it IDA}]{\includegraphics[width=0.3\linewidth]{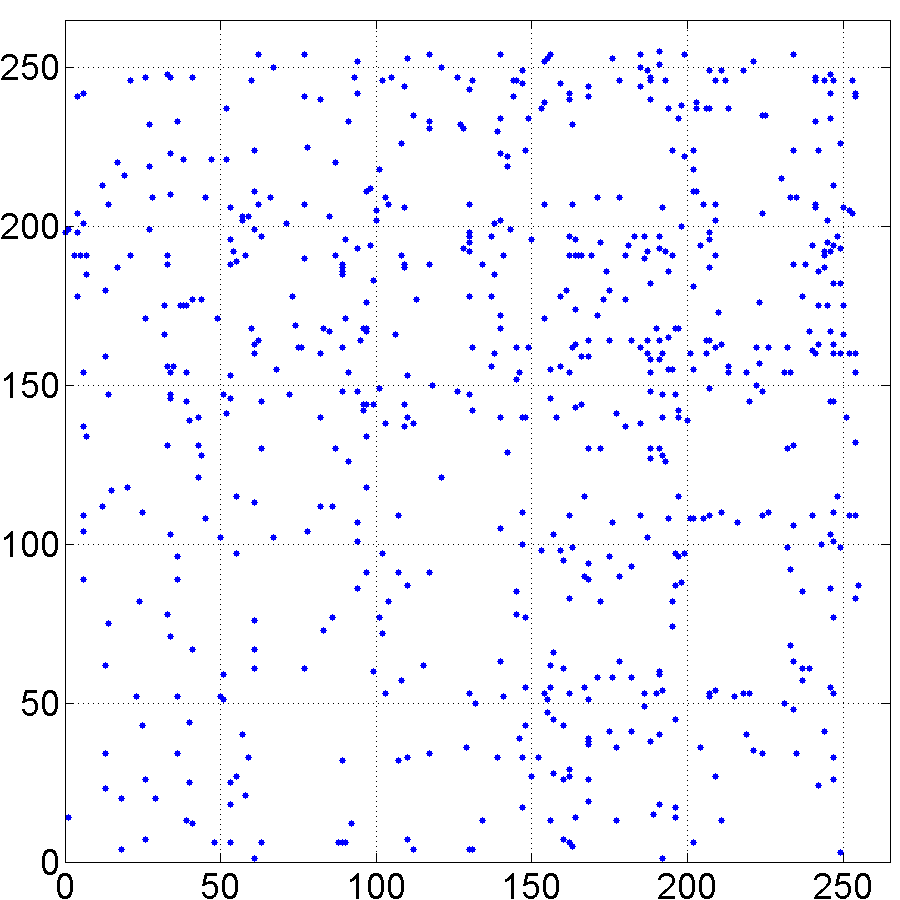}\label{fig:recf}}
\subfloat[][{\it Proposed scheme}]{\includegraphics[width=0.3\linewidth]{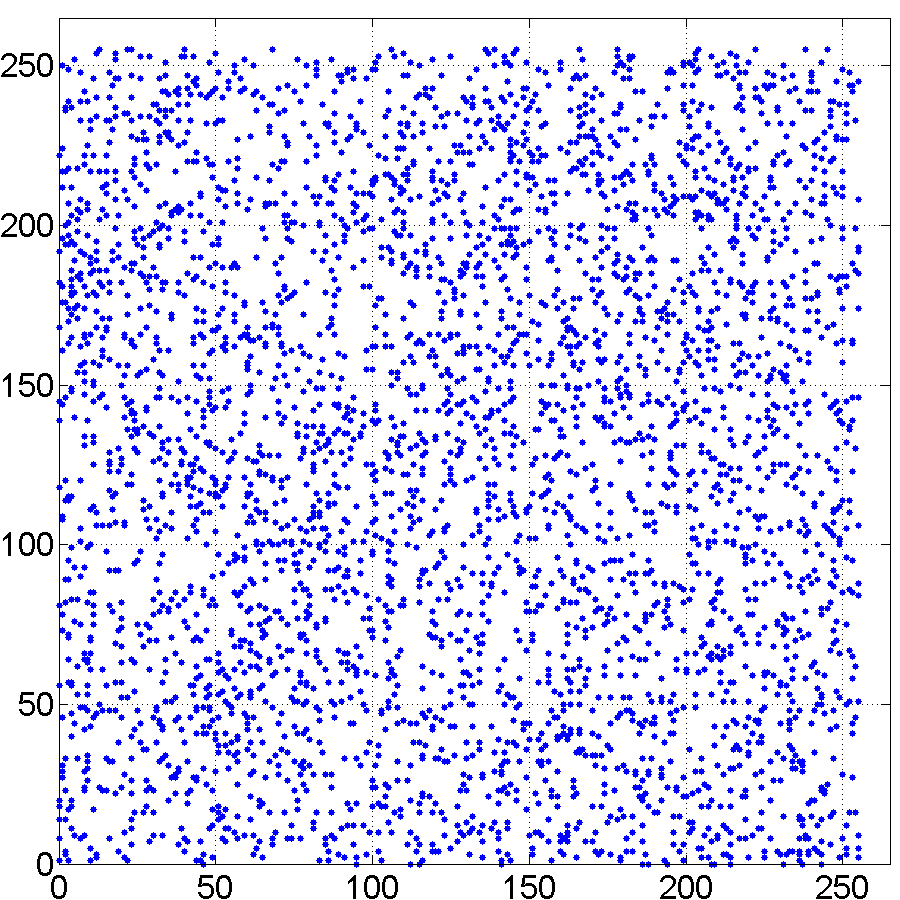}\label{fig:recf}}
\caption{{\it Recurrence plots for data from Figure~\ref{fig:distr}. x-axis shows data values, y-axis shows data values with a delay $t=1$. No clear pattern is obtained for a fragment obtained using the proposed scheme.}}%
\label{fig:REC}%
\end{figure}

A recurrence plot serves to estimate correlation inside data~\cite{correlation}. Considering data vector $x={x_1,x_2,...,x_m}$ a vector with delay $t \geq 1$ is constructed $x(t)=x_{1+t},x_{2+t},...,x_{m+t}$. A recurrence plot shows the variation between $x$ and $x(t)$. In Figure~\ref{fig:REC}, such plots for a data sample and its fragments obtained by applying an IDA and the proposed scheme are shown. Using the proposed scheme, no clear pattern is obtained after data fragmentation. 

\vspace{-1.5em}

\subsubsection{Correlation}

Correlation coefficient is used to evaluate the linear dependence between data~\cite{correlation}. A secure fragmentation algorithm should ensure as low correlation as possible between initial data and their fragments. 

\begin{figure}[!h]
\centering
\includegraphics[width=1\linewidth]{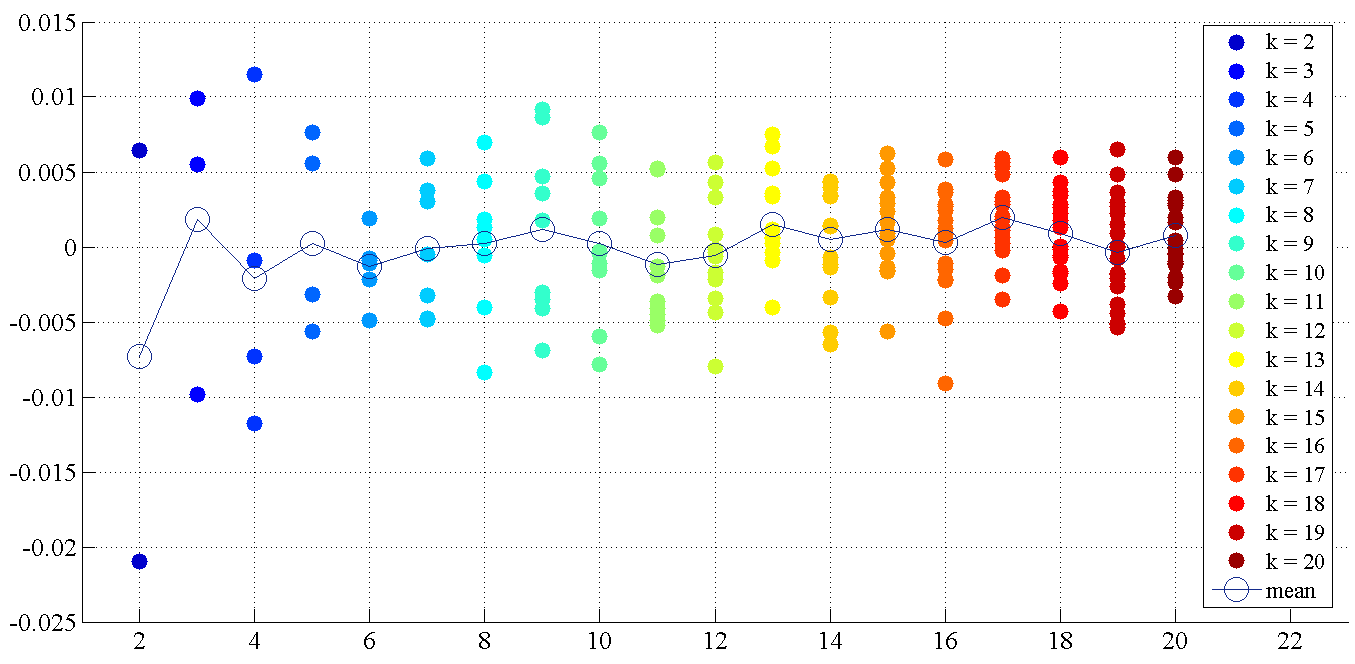}
\caption{{\it Correlation coefficients in function of $k$. x-axis shows the fragmentation identifier, y-axis shows the correlation coefficient between a data sample and its fragment. A correlation coefficient close to 0 indicates that there is no correlation between original data and its fragment. Black circles show the mean value of coefficients. With increasing of $k$ the mean values tend to be closer to 0.}}%
\label{fig:corrK}%
\end{figure}

Correlation coefficients were measured between 10 data samples and their corresponding fragments for different values of number of fragments $k$ (from 2 to 20). The method used for the calculation was same as in~\cite{coefficient:calcul}. Results are shown in Figure \ref{fig:corrK}. Observed values of correlation coefficients are close to 0. This demonstrates that no detectable correlation exists between tested data samples and their fragments. Moreover, the correlation coefficients for higher values of $k$ tend to have lower values. Then, the correlation between fragments coming from the same fragmentation results was measured. Correlation coefficients among fragments were also close to 0 (in a range of $<-0.01,0.02>$). It demonstrated that fragments are not correlated with each other and thus confirmed the independence property of the scheme. 

\subsubsection{Difference}

Each fragment should be significantly different from the initial data and from other fragments of the same fragmentation result. Bit difference between a data sample and each of its fragments was measured and it was close to 50\%. A similar result was obtained for the difference between fragments themselves.

\subsection{Sensitivity Test}
\label{ssec:seed-sensitivity}

Differential attacks study the relation between two ciphertexts resulting from a slight change, usually of one bit in the original plaintext or in the key.
Inspired by this fact, a seed sensitivity test was realized in order to visualize the impact of a change in the seed on the fragmentation result. Indeed, two fragmentation results of same data should be different while obtained using two distinct seeds. For same data sample of size of 4000 bytes, two fragmentation results were obtained: one for a seed $S_1$ and one for a seed $S_2$ that differs $S_1$ by one random bit. The test has shown that such two fragmentation results are not correlated with each other (correlation coefficients close to 0) and significantly different (around 50\% of bit difference).


\section{Cryptanalysis Discussion}
\label{sec:cryptanalysis-discussion}

In this Section, we discuss about the confidentiality level provided by the proposed algorithm ans its resistance against most known types of attack scenarios \cite{attack:brute,attack:differential,Paar:2009} (statistical, differential, chosen/known plain-text, and brute-force) in a situation when fewer than $k$ fragments have been revealed. During such study, the steps of the algorithm are considered to be public in compliance with Kerckhoff's principle.

\subsection{General Security Properties}
\label{ssec:properties}

The threat model we use is the one where data fragments are physically dispersed over $n$ different storage locations or transmission paths. Therefore, resulting data protection relies essentially on the difficulty to collect $k$ of the fragments. Indeed, an attacker has to find the locations or transmission paths of dispersed fragments and then manage to access or eavesdrop them. In this paper, we do not deal with data dispersal questions, which includes ensuring the physical separation of fragments, as well as protecting the information describing the order in which fragments have to be assembled to reconstitute the original data. Nevertheless information about the defragmenation procedure should be stored in a secure location or dispersed within the fragments, as its importance could be in a sense compared to the one of the encryption key.  


The strength of our solution relies on the use of a seed in form of $k$ pseudo-random values, as well as on unpredictability and high sensitivity of the fragments. Moreover, a user have the possibility to adapt the security level to their needs, not only by increasing the value of number of fragments $k$, but also by augmenting the data chunk size. The proposed scheme ensures forward and backward secrecy, but only in a situation when the seed values are not being repeated.

\subsection{Most known types of attacks}

In this Section we are analyzing the resistance of the scheme to the most known types of attacks.

\label{ssec:attacks}

{\bf Statistical attacks} The category of statistical attacks exploits the fact that encoded data may reveal some statistical properties. Immunity against such attacks requires that fragments achieve high level of randomness~\cite{uniformity}. Therefore, in an ideal situation, the frequency analysis of data inside a fragment should be indistinguishable from the output of a pseudorandom generator. Results presented in Section~\ref{sec:security} have shown that our scheme achieve good uniformity and recurrence characteristics. Indeed, no useful information can be detected from the fragmented data. This demonstrates the high randomness of the scheme and its resistance against statistical attacks. Such property is not achieved by Information Dispersal Algorithms, where in the case of a known generator matrix and data with recognizable patterns it is possible to guess the content of missing fragments. 

{\bf Brute-force attack and content verification} While possessing a $p<k$ number of fragments, an attacker can attempt a brute-force attack by trying to guess the content of the missing $(k-p)$ fragments. Intuitively, the difficulty level of such attack grows with the required number of fragments $k$ and decreases with the number of possessed fragments $p$. The recovery of a set of $k$ data shares of size $w$ each implicates trying $2^w$ possibilities for $k-p$ data shares of missing fragments. Therefore, an increasing of the size of a data chunk/share may harden the brute-force attack on a set of $k$ data chunks. On the other hand, a way of facilitating the attack would be to make some assumptions about the content of the missing fragments that would limit the number of possibilities to verify.

In a different scenario, an attacker with less than $k$ fragments may like to verify whether the data inside fragments matches some predetermined value.
In the case of an IDA, such verification is easy when the generator matrix is known. In the case of our scheme, it is harder, as the attacker will have to guess or the missing seed values or the missing data shares used to encode the part of the data they would like to verify.

{\bf Known and chosen plain text attacks} The knowledge of a part of the plaintext facilitates a brute-force attack. However, as each time a new seed is used, it does not help to recover fragments of other data.

{\bf Linear and differential attacks} A linear attack consists of exploiting the linear relations between the plaintext, the ciphertext and the key. The knowledge of the first data chunks from the plaintext and of the $p$ fragments allows the recovery of the missing seed values. However, the recovery of the seed is not as critical, as the key leakage in symmetric encryption schemes. Because fragments are dependent of each others, to decode a  data share it is necessary to possess its preceding data shares, also the one that are inside of the $k-p$ missing fragments. 
In order to avoid differential attacks, seed values have to be change for each fragmentation procedure. A reuse of the seed could expose some relations between fragmentation results, for instance the fragmentation result of two identical plaintexts would be the same. As presented in Section ~\ref{ssec:seed-sensitivity}, encoded data show high seed sensitivity, so even a single bit change is sufficient to obtain two different and not correlated fragmentation results of the same plain text.

\section{Performance Results}
\label{sec:performance}

\begin{figure}[!h]
\centering
\includegraphics[width=1\linewidth]{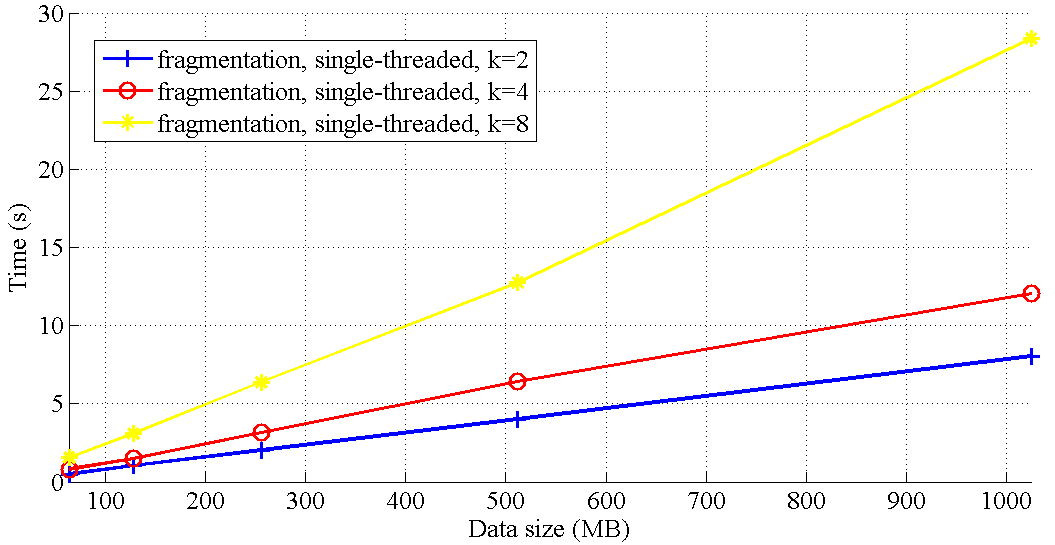}
\caption{{\it Performance measured on a portable computer. Variation of the average fragmentation time, single-threaded implementation, k = 2, 6, 8.}}
\label{fig:laptop}%
\end{figure}

Proposed algorithm was implemented in JAVA using the following resources:  JDK 1.8 on DELL Latitude E6540, X64-based PC running on Intel\textsuperscript{\textregistered} Core\textsuperscript{TM} i7-4800MQ CPU @ 2.70 GHz with 8 GB RAM, under Windows 7. It was tested on data provided by La Poste, the French postal office. An implementation of $java.security.SecureRandom$ is used for seed generation. The scheme can be implemented in any $GF(2^Q)$ and is designed to use only logical operations. For integrity purposes, $Q$ is selected according to word size of processors  and can be 8,16, 32 or 64-bit. As presented in Figure~\ref{fig:laptop}, the variation of average time for fragmentation is linear. Similar results were obtained for the defragmentation process. A multi-threaded version, optimized for 4 cores, sped up the performance by a factor of 3 that becomes close to 4 for more intensive computations.

\begin{figure}[!h]
\centering
\includegraphics[width=1\linewidth]{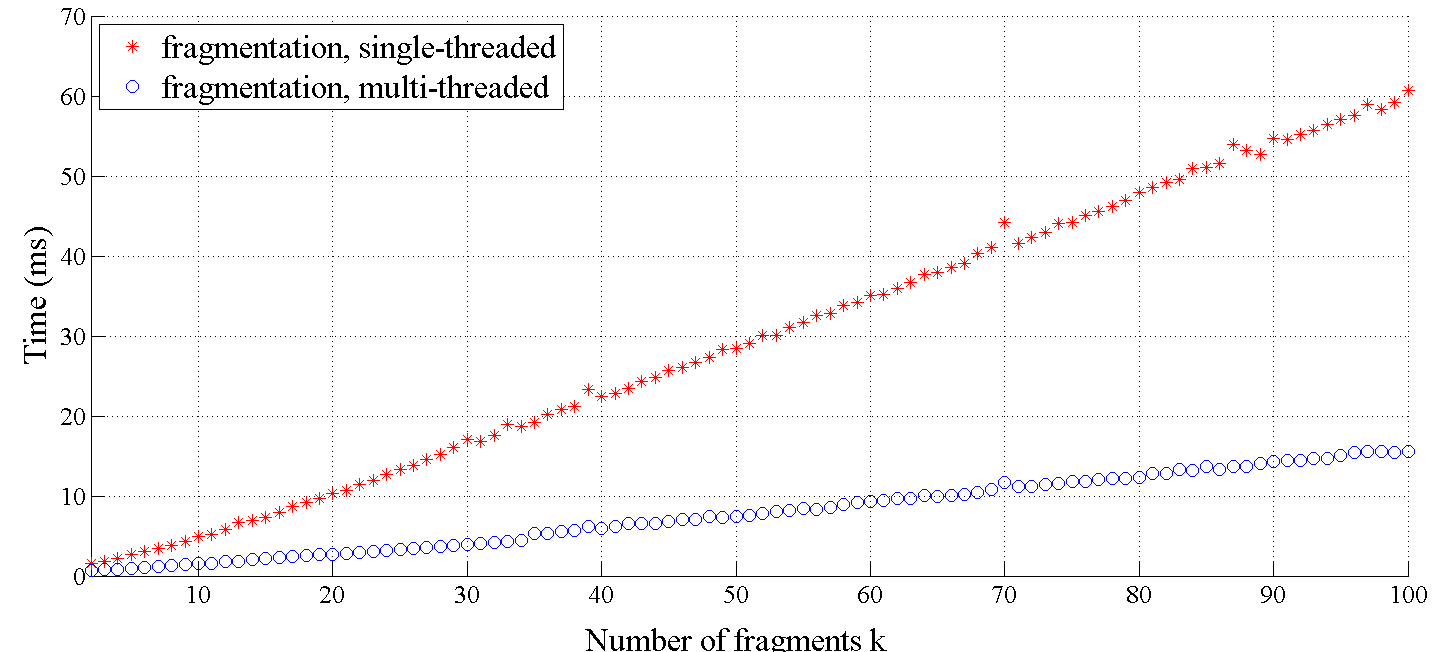}
\caption{{\it Performance measured on the TeraLab platform. Variation of the average fragmentation time for values of $k$ from 2 to 100. Data sample size is equal to 100MB. The multi-threaded version of the code was optimized to use 4 threads. }}
\label{fig:perf100}%
\end{figure}

\begin{figure}[!h]
\centering
\includegraphics[width=1\linewidth]{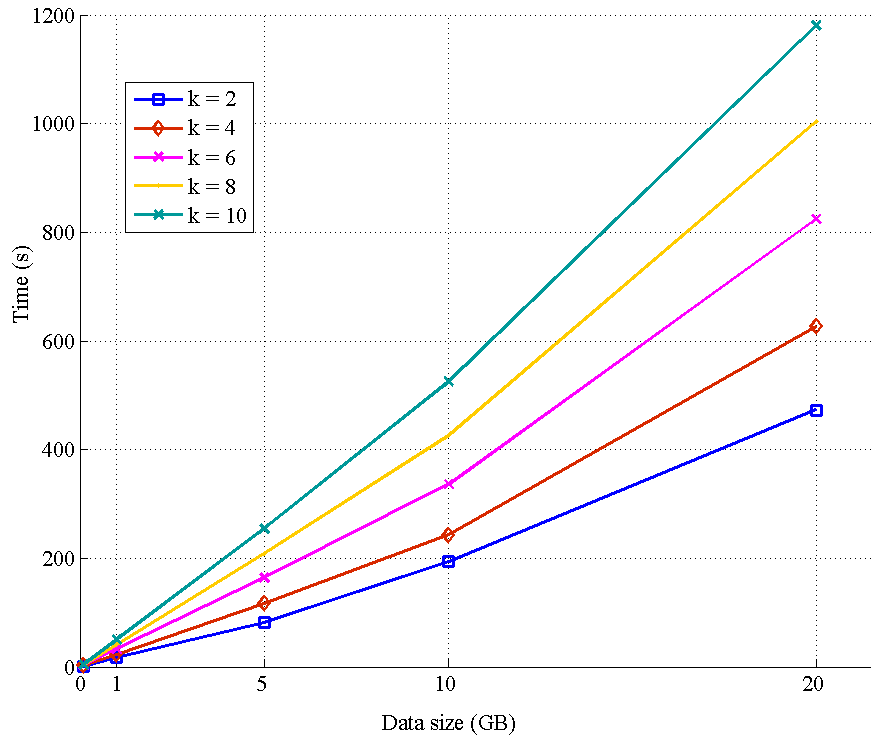}
\caption{{\it Performance measured on the TeraLab platform. Variation of the average fragmentation time in function of data size for the single-threaded version of the application for a data size up to 20GB. }}
\label{fig:bigdata}%
\end{figure}

Another test was performed on the tera-memory platform TeraLab\footnote{https://www.teralab-datascience.fr} using following resources: 32GB RAM and 4 VCPUs. Figure~\ref{fig:perf100} shows the results for fragmentation of a data file of 100MB up to 100 fragments. This result exhibits the linear complexity of the fragmentation relatively to $k$ the number of generated fragments. Figure~\ref{fig:bigdata} presents performance measured in function of data size for several values of $k$. Similar results were obtained for the defragmentation process. Results demonstrate that the scheme achieves good scalability.

\section{Future Works}
\label{sec:future}

In future, we plan to benchmark our algorithm to compare its performance with most relevant works. A possibility of performance improvement is seen in parallelization of the fragmentation processing by partially limiting the sequential character of the encoding procedure. A more sophisticated way of distribution of encoded data chunks could be also envisioned in order to create fragments taking into account several levels of confidentiality. Such processing would require an additional step in form of the separation of initial data chunks regarding their level of confidentiality. It could be inspired by works done in \cite{Qiu:2015}, that use wavelet transform to separate data into two parts, a private and a public one, without any user interaction.

Another, more complex, research track would be to design a complete fragmentation architecture for data protection by means of fragmentation, encryption, and dispersion. This work would focus on secure management of information about the location and order of data fragments, as well as on secure distribution of fragments from the trusted site performing fragmentation procedure to their final storage destinations.

\section{Conclusion}
\label{sec:conclusion}

A novel fragmentation algorithm for secure and resilient distributed data storage was described and analyzed. It combines the keyless property, computational simplicity and space efficient size of fragment of Information Dispersal Algorithms with an adjustable computational level of security. Security analysis shows that the produced fragments achieve good randomness and are not correlated neither with the initial data, nor among themselves.                                                                                  
The use of a fresh seed for each fragmentation procedure ensures the backward and forward secrecy properties, as well as limits the possibility of differential attacks. Brute-force and chosen-plaintext attacks could be considered in some situations. Their attack resistance depends on the number of fragments in the possession of an attacker, as well as the quantity of knowledge about of the initial data.

The scheme was implemented and tested on different samples of industrial data. Tests show good performance and scalability. The fragmentation procedure is linear in terms of number of fragments. 

We believe that the scheme could be applied to all application of data storage or transmission, where we would like to hide the nature of fragmented data without applying cryptographic mechanisms.

\section{Acknowledgments}

The work is partially funded by the ITEA2 CAP project.

\bibliographystyle{abbrv}
\bibliography{bibliography}

\begin{thebibliography}{10}

\bibitem{cuppens}
A.~Bkakria, F.~Cuppens, N.~Cuppens-Boulahia, and J.~M. Fernandez.
\newblock Confidentiality-preserving query execution of fragmented outsourced
  data.
\newblock In {\em Proceedings of the 2013 International Conference on
  Information and Communication Technology}, ICT-EurAsia'13, pages 426--440,
  Berlin, Heidelberg, 2013. Springer-Verlag.

\bibitem{secret:shares}
W.~J. Buchanan, D.~Lanc, E.~Ukwandu, L.~Fan, G.~Russell, and O.~Lo.
\newblock The future internet: A world of secret shares.
\newblock {\em Future Internet}, 7(4):445, 2015.

\bibitem{entropy}
C.~Cachin.
\newblock Entropy measures and unconditional security in cryptography, 1997.

\bibitem{attack:brute}
J.-S. Cho, S.-S. Yeo, and S.~K. Kim.
\newblock Securing against brute-force attack: A hash-based rfid mutual
  authentication protocol using a secret value.
\newblock {\em Comput. Commun.}, 34(3):391--397, Mar. 2011.

\bibitem{systemx}
P.~Cincilla, A.~Boudguiga, M.~Hadji, and A.~Kaiser.
\newblock Light blind: Why encrypt if you can share?
\newblock In {\em 2015 12th International Joint Conference on e-Business and
  Telecommunications (ICETE)}, volume~04, pages 361--368, July 2015.

\bibitem{chi2}
W.~G. Cochran.
\newblock The $\chi^2$ test of goodness of fit.
\newblock {\em Ann. Math. Statist.}, 23(3):315--345, 09 1952.

\bibitem{bib:vimercati}
S.~De~Capitani~di Vimercati, R.~F. Erbacher, S.~Foresti, S.~Jajodia,
  G.~Livraga, and P.~Samarati.
\newblock {\em Encryption and Fragmentation for Data Confidentiality in the
  Cloud}, pages 212--243.
\newblock Springer International Publishing, Cham, 2014.

\bibitem{Dworkin:2001:SER:2206247}
M.~J. Dworkin.
\newblock Sp 800-38a 2001 edition. recommendation for block cipher modes of
  operation: Methods and techniques.
\newblock Technical report, Gaithersburg, MD, United States, 2001.

\bibitem{Fabre1994}
J.-C. Fabre, Y.~Deswarte, and B.~Randell.
\newblock {\em Designing secure and reliable applications using
  fragmentation-redundancy-scattering: an object-oriented approach}, pages
  21--38.
\newblock Springer Berlin Heidelberg, Berlin, Heidelberg, 1994.

\bibitem{coefficient:calcul}
Z.~Fawaz, H.~Noura, and A.~Mostefaoui.
\newblock An efficient and secure cipher scheme for images confidentiality
  preservation.
\newblock {\em Image Commun.}, 42(C):90--108, Mar. 2016.

\bibitem{Kapusta:2016}
K.~Kapusta, G.~Memmi, and H.~Noura.
\newblock Poster: A keyless efficient algorithm for data protection by means of
  fragmentation.
\newblock In {\em Proceedings of the 2016 ACM SIGSAC Conference on Computer and
  Communications Security}, CCS '16, pages 1745--1747, New York, NY, USA, 2016.
  ACM.

\bibitem{Katti:slicing}
S.~Katti, J.~Cohen, and D.~Katabi.
\newblock Information slicing: Anonymity using unreliable overlays.
\newblock In {\em Proceedings of the 4th USENIX Conference on Networked Systems
  Design \&\#38; Implementation}, NSDI'07, pages 4--4, Berkeley, CA, USA, 2007.
  USENIX Association.

\bibitem{Krawczyk:1993:SSM:646758.705700}
H.~Krawczyk.
\newblock Secret sharing made short.
\newblock In {\em Proceedings of the 13th Annual International Cryptology
  Conference on Advances in Cryptology}, CRYPTO '93, pages 136--146, London,
  UK, 1994. Springer-Verlag.

\bibitem{Li:ida}
M.~Li.
\newblock On the confidentiality of information dispersal algorithms and their
  erasure codes.
\newblock {\em CoRR}, abs/1206.4123, 2012.

\bibitem{noura}
H.~Noura, S.~Martin, K.~A. Agha, and K.~Chahine.
\newblock Erss-rlnc: Efficient and robust secure scheme for random linear
  network coding.
\newblock {\em Computer Networks}, 75, Part A:99 -- 112, 2014.

\bibitem{attack:differential}
K.~Nyberg and L.~R. Knudsen.
\newblock {\em Provable Security Against Differential Cryptanalysis}, pages
  566--574.
\newblock Springer Berlin Heidelberg, Berlin, Heidelberg, 1993.

\bibitem{Paar:2009}
C.~Paar and J.~Pelzl.
\newblock {\em Understanding Cryptography: A Textbook for Students and
  Practitioners}.
\newblock Springer Publishing Company, Incorporated, 1st edition, 2009.

\bibitem{Parakh:onlinedatastorage}
A.~Parakh and S.~Kak.
\newblock Online data storage using implicit security.
\newblock {\em Inf. Sci.}, 179(19):3323--3331, Sept. 2009.

\bibitem{Parakh:2011335}
A.~Parakh and S.~Kak.
\newblock Space efficient secret sharing for implicit data security.
\newblock {\em Information Sciences}, 181(2):335 -- 341, 2011.

\bibitem{Qiu:2015}
H.~Qiu and G.~Memmi.
\newblock Fast selective encryption methods for bitmap images.
\newblock {\em Int. J. Multimed. Data Eng. Manag.}, 6(3):51--69, July 2015.

\bibitem{rabin:1989}
M.~O. Rabin.
\newblock Efficient dispersal of information for security, load balancing, and
  fault tolerance.
\newblock {\em J. ACM}, 36(2):335--348, Apr. 1989.

\bibitem{RS}
I.~S. Reed and G.~Solomon.
\newblock Polynomial {C}odes {O}ver {C}ertain {F}inite {F}ields.
\newblock {\em Journal of the Society for Industrial and Applied Mathematics},
  8(2):300--304, 1960.

\bibitem{Resch:2011}
J.~K. Resch and J.~S. Plank.
\newblock Aont-rs: Blending security and performance in dispersed storage
  systems.
\newblock In {\em Proceedings of the 9th USENIX Conference on File and Stroage
  Technologies}, FAST'11, pages 14--14, Berkeley, CA, USA, 2011. USENIX
  Association.

\bibitem{Rivest:97}
R.~L. Rivest.
\newblock All-or-nothing encryption and the package transform.
\newblock In {\em In Fast Software Encryption, LNCS}, pages 210--218.
  Springer-Verlag, 1997.

\bibitem{correlation}
J.~L. Rodgers and A.~W. Nicewander.
\newblock {Thirteen Ways to Look at the Correlation Coefficient}.
\newblock {\em The American Statistician}, 42(1):59--66, 1988.

\bibitem{Shamir:1979:SS:359168.359176}
A.~Shamir.
\newblock How to share a secret.
\newblock {\em Commun. ACM}, 22(11):612--613, Nov. 1979.

\bibitem{uniformity}
S.~Xu, Y.~Wang, J.~Wang, and M.~Tian.
\newblock Cryptanalysis of two chaotic image encryption schemes based on
  permutation and xor operations.
\newblock In {\em Computational Intelligence and Security, 2008. CIS '08.
  International Conference on}, volume~2, pages 433--437, Dec 2008.

\end{thebibliography}

\end{document}